\documentclass[letterpaper, aps, tightenlines, nofootinbib, 11pt]{revtex4}

\pdfoutput=1

\usepackage[pdftex, bookmarks = false, colorlinks = true, linkcolor = blue, citecolor = purple]{hyperref}

\usepackage{shortcuts}

\usepackage{titlesec}

\def\thesection{\arabic{section}} 
\def\thesubsection{\arabic{subsection}} 
\titleformat{\section}
  {\normalfont\Large\bfseries}{\thesection}{1em}{}
\titleformat{\subsection}
  {\normalfont\large\bfseries}{\thesubsection}{1em}{}
\titleformat{\subsubsection}
  {\normalfont\it}{\thesubsection.\thesubsubsection}{1em}{}

\preprint{\tt PUPT-2379}

\begin{document}

\title{\Large Connecting the Holographic and Wilsonian\\ Renormalization Groups\\ }
\author{\large \DJ or\dj e Radi\v cevi\'c\\ }
\affiliation{ Department of Physics, Princeton University,\\ Princeton, NJ 08544, USA\\ \\ \rm }
\affiliation{ \it Department of Physics, Stanford University,\\ Stanford, CA 94305, USA \\ \\ {\tt djordje@stanford.edu}\\ }

\begin{abstract}
\noindent Inspired by the AdS/CFT duality, we develop an explicit formal correspondence between the planar limit of a $d$-dimensional global gauge theory and a classical field theory in a $(d + 1)$-dimensional anti-de Sitter space. The key ingredient is the identification of scalar fields in the AdS with generalized Hubbard-Stratonovich transforms of single-trace couplings of the QFT. Guided by this idea, we show that the Wilsonian renormalization group flow of these transformed couplings can match the holographic (Hamilton-Jacobi) flow of bulk fields along the radial direction in AdS. This result leads to an outline of an AdS/CFT dictionary that does not rely on  string theory.
\end{abstract}

\maketitle

\section{Introduction}

It has long been understood that the AdS/CFT correspondence \cite{Maldacena, GKP, Witten} hints at an effective mapping between a gravity theory in a $(d + 1)$-dimensional bulk and a gauge theory in a flat $d$-dimensional space at the boundary of the bulk. A considerable research effort has been devoted to studying the phenomenology of such a duality. Many remarkable similarities between different bulk and boundary theories are now known; reviews and references can be found in \cite{Policastro, Hartnoll, HerzogLectures, McGreevy}. The progress achieved in this direction naturally prompts us to ask whether a bulk theory can be shown to be dual to a boundary theory without even invoking the string-theoretic roots of AdS/CFT. 

A pivotal role in these considerations is played by the notion that the bulk dynamics should be closely related to Wilsonian RG flows of boundary couplings; e.g.~see \cite{Skenderis} and references therein. In particular, it is known that the AdS/CFT dictionary maps Callan-Symanzik equations for a boundary QFT onto Hamilton-Jacobi (``holographic RG'') equations in an anti-de Sitter (AdS) bulk \cite{deBVV}. It was also suggested \cite{Khoury, VV, Li, Akhmedov2, Akhmedov3} that the holographic RG could be explicitly related to Polchinski's ``exact'' formulation of the Wilsonian RG  \cite{Polchinski}. Very recently, these ideas echoed through a series of works commenting on relations between different aspects of holographic and Wilsonian RGs \cite{Heemskerk2, Faulkner, Douglas, Strominger}.  Nevertheless, we still lack a precise statement on how to reach the bulk dual of a QFT by computing its Wilsonian RG flow.

A different path towards this goal was recently opened when it was shown that one can recast a $d$-dimensional gauge theory into a $(d + 1)$-dimensional field theory, with bulk fields corresponding to Hubbard-Stratonovich transforms of boundary couplings, and with the extra dimension corresponding to the energy scale of the boundary theory \cite{Lee1}. It was also shown that one can similarly connect a planar (large $N$) $d$-dimensional lattice gauge theory with a field theory on a lattice version of AdS$_{d + 1}$ \cite{Lee2}. This ties in well with the AdS/CFT claim that the large $N$ limit on the boundary should correspond to the classical limit in the bulk. These results hint that an effective gauge/gravity duality may perhaps link bulk fields and certain transforms of boundary couplings.

As we show in this paper, the analogy between holographic and Wilsonian RGs, and the corresponding need for some type of a Hubbard-Stratonovich transformation, can arise even without appealing to AdS/CFT. In general, if we can describe the same physics using field theories with different numbers of degrees of freedom, then we can formulate an RG flow in the space of theories by integrating out degrees of freedom and reabsorbing them into the parameters of the theories. Working within this framework, we provide an example of a particular bulk theory whose holographic RG flow is formally equivalent to the Wilsonian RG flow of a particular boundary theory.

In the context of Wilsonian RGs, we describe the physics using matrix-valued quantum fields whose high energy modes are integrated out and whose couplings are set by this integration \cite{Wilson1, Wilson2}. This integration can be done step-by-step. If we take a theory that is cut off at some cutoff $\Lambda$ and integrate out modes in the momentum shell at $\Lambda$, we get a new theory which has the same partition function as the old one --- but which also has a different set of couplings and a momentum cutoff at $\Lambda - \d\Lambda$. The Wilsonian RG equations govern how couplings change as $\Lambda$ is lowered during this procedure.

As for holographic RGs, if we take our bulk to be an AdS space with radial coordinate $z$, the dynamics of bulk fields at $z \geq \eps$ is specified by the bulk action \emph{and} by the boundary conditions (values of fields at $z = \eps$). The latter are set by integrating out fields at $z < \eps$. This integration can be done piecemeal, just like in the Wilsonian case. If we want to move the boundary to $z = \eps + \d\eps$, we integrate out the bulk fields in the AdS slice near $z = \eps$ and we find how boundary conditions change as we go from one boundary to the other. This gives the  holographic RG equations.

In the limit of large $N$ (representing one type of a classical limit in the boundary), both RGs boil down to the classical Hamilton-Jacobi (HJ) theory \cite{Landau}: an action $S$ describes the physics in some region with a boundary at position $t$, and shifting this boundary results in the flow of some parameters $\phi$ (couplings or boundary conditions) needed to specify the physics inside the region. This flow is captured by an HJ equation for an appropriate functional $I[\phi; t]$:
\bel{\label{HJ}
  \pder{I[\phi; t]}{t} = H\left(\phi, \pi, t\right),\quad \pi \equiv \pder{I[\phi; t]}\phi.
}
Here $H$ is the Hamiltonian function of the system and $\pi$ are the momenta conjugate to $\phi$. 

Asking for the meaning of ``appropriate $I[\phi; t]$'' reveals the crucial difference between the two RGs. The holographic RG has a simple structure: for a bulk action $S[\Phi; \eps]$, the fields $\Phi$ both flow and get integrated out, and the $I$ entering (\ref{HJ}) is just the on-shell value of $S$ which depends on $\eps$ and on the boundary fields $\phi \equiv \Phi|_{z = \eps}$. On the other hand, for a theory with matrix fields $M$ and action $S[M; \Lambda]$, fields $M$ are being integrated out while couplings $\phi$ flow, and it is unclear what $I$ should be. As we show, this flow can be recast into an HJ equation using a generalized Hubbard-Stratonovich transformation of the boundary action, which we define as
\bel{\label{HST preview}
  e^{-S[M; \Lambda]} \equiv \int [\d \varphi] e^{-N \int \varphi_n  \Tr M^n  - N^2 \~S[\varphi; \Lambda]}.
}
This transformation transfers the burden of ``having to be integrated out'' from quantum fields $M$ to couplings $\phi$; the cutoff dependence is transfered from the Wilsonian action $S[M; \Lambda]$ to an action functional $\~S[\varphi; \Lambda]$ which does \emph{not} depend on the quantum fields $M$. The transformed action $\~S[\varphi; \Lambda]$ is now the natural candidate for $I$ in eq.~(\ref{HJ}). Indeed, our main result is that the Wilsonian RG flow for $\~S[\varphi; \Lambda]$  can (under certain conditions) take the form of the \emph{same} HJ equation that encodes the holographic RG flow for fields in the AdS.

We will make this statement more precise in the remainder of this paper. We will compute the RG flows for a simple bulk-boundary pair whose choice is loosely inspired by the AdS/CFT dictionary.  (We will \emph{not}, however, make any assumption that AdS/CFT holds.) On the bulk side (Section \ref{sec HRG}), we will find the holographic RG equations which describe the classical dynamics of a set of interacting scalar fields in an AdS space. On the boundary side (Section \ref{sec WRG}), we will consider a Euclidean QFT of a single matrix field $M$ with a global $O(N)$ symmetry. We will show how the RG flow of all couplings can be elegantly captured by Polchinski's exact RG equation, and then we will perform the generalized Hubbard-Stratonovich transformation (\ref{HST preview}) of $S[M; \Lambda]$.
Using the large $N$ limit, we will then be able to write the Wilsonian RG equation for $\~S[\varphi; \Lambda]$ in the same form as the holographic RG equation from Section \ref{sec HRG}; both will take the form of eq.~(\ref{HJ}). The final result will thus be an explicit demonstration that, under the right conditions, Wilsonian RG flows in a $d$-dimensional, Euclidean, planar theory with global $O(N)$ symmetry display the same dynamics as holographic RG flows of a classical field theory in AdS$_{d + 1}$.

\section{The holographic RG} \label{sec HRG}

Our bulk will be a simple $(d + 1)$-dimensional space with a fixed anti-de Sitter metric. We will neglect the back-reaction of the metric. The line element is
\bel{
  \d s^2 = g_{\mu\nu} \d x^\mu \d x^\nu = \frac{R^2}{z^2}(\d z^2 + \delta_{ij} \d x^i \d x^j).
}
The coordinate $x^d \equiv z$ refers to the ``radial direction'' in the AdS, $R$ is the constant AdS radius, and it is assumed that a $z=$ const slice has Euclidean metric. The metric diverges at $z = 0$, but we will consider only regions at $z > 0$ and so this singularity will not affect us. We consider an action for many scalar fields propagating at $z \geq \eps$:
\algnl{\label{def S AdS}
  S\_{AdS}[\Phi; \eps] 
  &= \int_{z \geq \eps}\!\! \d^{d + 1}x \sqrt g \left[\frac12 G^{nm} \del_\mu \Phi_n \del^\mu \Phi_m + V\_{AdS}(\Phi)\right].
}  
We will assume that $G_{nm}$ and $V\_{AdS}$ are analytic functions of $z$. 

The appropriate HJ equation is easy to find. Thinking of $z$ as ``time'' and using the explicit form for the metric, the Lagrangian of this scalar theory is
\algnl{
  L\_{AdS}(\Phi, \del_z \Phi, z) = \int \d^d x\ \frac{R^{d + 1}}{z^{d + 1}} \left[\frac12 G^{nm} \frac{z^2}{R^2}\del_z \Phi_n \del_z \Phi_m + \left(\frac12 G^{nm} \frac{z^2}{R^2} \del_i \Phi_n \del^i \Phi_m  + V\_{AdS}(\Phi)\right)\right].
}
Let $h_{ij} \equiv (R/z)^2 \delta_{ij}$ be the induced metric on a constant $z$ slice. The natural conjugate momenta are
\bel{
  \Pi^n \equiv \frac1{\sqrt{h}} \fder{L\_{AdS}}{(\del_z\Phi_n)} = \frac z R G^{nm}\del_z \Phi_m.}
The Hamiltonian is then
\bel{
  H\_{AdS}(\Phi, \Pi, z) = \int \d^d x\ \sqrt h\ \Pi^n \del_z \Phi_n - L\_{AdS} = \frac R z\int \d^d x\ \sqrt h \left[\frac12 G_{nm} \Pi^n \Pi^m - V(\Phi, z)\right],
}
and the potential is
\bel{\label{V}
  V(\Phi, z) =  \frac12 \frac{z^2}{R^2} G^{nm} \del_i\Phi_n \del^i \Phi_m + V\_{AdS}(\Phi).
}
Knowing the Hamiltonian we can write down the HJ equation (\ref{HJ}) for $I$, the on-shell value of $S\_{AdS}$:
\bel{\label{flow}
  \eps \pder{I[\phi; \eps]}\eps = R \int \d^d x \ \sqrt h  \left[\frac12 G_{nm} \pi^n\pi^m - V(\phi, \eps)\right],
}
with endpoint fields and momenta 
\bel{
  \phi_n \equiv \Phi_n\big|_{z = \eps}, \quad \pi^n \equiv \Pi^n\big|_{z = \eps} = \frac1{\sqrt h} \fder I{\phi_n}\biggr|_{z = \eps}.
}

Note that the partition function of the bulk theory will be fully determined at classical level, i.e.~in the saddle point limit, by solving the HJ equation for $I$. Incidentally, if we were interested in applying the standard AdS/CFT dictionary, we would want to solve precisely for $I[\phi; \eps]$. This functional (along with extra boundary counterterms which we will not discuss here) then gives the partition function $Z = e^{-I}$, which can then be differentiated w.r.t.~the sources $\phi$ to obtain the correlation functions in the boundary.

\section{The Wilsonian RG} \label{sec WRG}

It remains to show how a Wilsonian RG flow can be reduced to an equation like (\ref{flow}). The AdS/CFT dictionary standardly suggests that we should work with an $SU(N)$ gauge theory with large $N$ and  near an RG fixed point. We will choose a simple model roughly along those lines: a theory of a single symmetric $N \times N$ matrix field $M(x)$ with real entries. We will realize the near-conformality assumption by assuming that all couplings are sufficiently small so that the theory is near a free-field fixed point; this is a weaker assumption than near-conformality, but it is all we need. We will also impose a global $O(N)$ symmetry, so the general action will feature only traces of matrices:
\algnl{\notag
  S[M; \Lambda] 
  &= \int \xdvol\ \left[N \phi_n \Tr M^n + \phi_{nm} \Tr M^n \Tr M^m + \trm{other\ multiple\ traces} \right]\\ \label{def S}
  &\equiv N \phi_n \cdot \Tr M^n + \Tr M^n \cdot \phi_{nm} \cdot \Tr M^m + \ldots
}
The couplings $\phi$ may contain finitely many derivatives. Notice that we denote integration over all space with a dot product invariant under Fourier transforms. This way we avoid cluttering our equations from having to transform back and forth between momentum and position space. Going to momentum space will be necessary in order to perform the RG step and to talk about couplings that contain derivatives, but we will want to express the end-result in position space.

It is of note that we impose a global $O(N)$ symmetry in the boundary theory, whereas many papers on the AdS/CFT correspondence assume a boundary theory which is locally gauge invariant. In addition, typical examples of AdS/CFT have strongly coupled boundary theories, whereas here we explicitly assume weak couplings. Once again, we stress that the theories whose RG equations we match will not be any conventional example of an AdS/CFT pair. One of the novel results of this paper will be to precisely exhibit how close the correspondence we develop approaches AdS/CFT.

The action (\ref{def S}) is written down using the standard $N$-counting conventions \cite{tHooft}: the single-trace term has a prefactor of $N$, and products of multiple traces enter the action with a suppression factor of $1/N$ for each additional trace in the multi-trace term. This $N$-counting remains consistent through the RG flow, and at large $N$ this makes sure we are left with a classical theory of single-trace operators. In addition, such $N$-counting is a step towards formulating a string theory from an $O(N)$ gauge theory, but we defer studying this connection to future work. It is important to emphasize that we will assume that $N$ is large throughout this paper. Most formulas will only involve leading terms in $1/N$ expansions.

We also assume that the theory has a cutoff $\Lambda$ in momentum space. We implement this by adding a factor of $K^{-1}_\Lambda(p^2)$ to the Fourier transform of the quadratic term $\Tr M^2$ in the action. The function $K_\Lambda(p^2)$ is chosen to be unity at $p^2 < \Lambda^2$ and zero otherwise, with a smooth but exponentially steep drop at $p^2 = \Lambda^2$. This means that the free-field propagator is
\bel{
  \avg{M_{ab}(p) M_{cd}(-p)} = K_\Lambda(p^2)\Delta(p)\delta_{ad}\delta_{bc} \quad\text{with}\quad \Delta(p) = \frac1{2N\phi_2(p)},
}
so high momentum modes do not propagate. Incidentally, we are already using the large $N$ limit here: we are ignoring corrections to the propagator coming from the other quadratic term, $\phi_{11} (\Tr M)^2$. This needs to be considered if one wants to compute higher-order corrections in $1/N$. 

As already discussed, the Wilsonian RG equation describes how the couplings $\phi$ flow as we integrate modes of the field $M$ in the momentum shell at $\Lambda$. The Polchinski equation (PE) gives us an efficient way of keeping track of all the couplings at the same time, as it describes the RG flow of the entire action $S[M; \Lambda]$ \cite{Polchinski}. Originally, the PE was derived only for scalar fields and for the interaction part of the action. To derive the matrix field analogue for the entire action, we generalize the practical derivation in \cite{Zinn-Justin}. First, we split the field into slow and fast modes, so that the fast modes only propagate at momenta in the momentum shell at $\Lambda$. Second, we expand the action in powers of the fast modes; this  will effectively be an expansion in powers of the small ratio $\d\Lambda/\Lambda$. Third, working to second order in $\d\Lambda/\Lambda$, we path-integrate these modes out in the partition function and express the result as the exponential of a new action. Finally, comparing this new action to the old one, we read off the PE for matrix fields
\vspace{2mm}
\bel{\label{PE}
  \dlam e^{-S[M; \Lambda]} = \op H_\Lambda e^{-S[M; \Lambda]},
}
where the differential operators $\dlam$ and $\op H_\Lambda$ are defined as 
\algnl{
  \dlam f(\Lambda) &\equiv f(\Lambda) - f(\Lambda - \d\Lambda),\\
  \op H_\Lambda &\equiv - \frac12 \int \pdnvol\ \dlam K_\Lambda(p^2) \Delta(p)  \fder{^2}{M_{ij}(p)\delta M_{ji}(-p)}.
}

\begin{figure}[b]
\begin{center}
\begin{picture}(50, 15)
  \put(-5, 4.5){$\dlam$}
  \put(-3, 4.5){$\Biggr($}
  \put(11.5, 4.5){$\Biggr)\ =$}
  \put(4, 5){\line(-1, -1){5}}
  \put(4, 5){\line(-1, 1){5}}
  \put(5, 6){\line(-1, 1){5}}
  \put(5, 6){\line(1, 1){5}}
  \put(6, 5){\line(1, 1){5}}
  \put(6, 5){\line(1, -1){5}}
  \put(5, 4){\line(1, -1){5}}
  \put(5, 4){\line(-1, -1){5}}
  \put(4, 8){$\phi_4$}

  \put(23, 5){\line(-1, 1){5}}
  \put(23, 5){\line(-1, -1){5}}
  \put(24, 5.75){\line(-1, 1){5}}
  \put(24, 4.25){\line(-1, -1){5}}
  {\color{red} \linethickness{1.5 pt}\put(24, 4.25){\line(1, 0){3}} \put(24, 5.75){\line(1, 0){3}}}
  \put(27, 5.75){\line(1, 1){5}}
  \put(27, 4.25){\line(1, -1){5}}
  \put(28, 5){\line(1, 1){5}}
  \put(28, 5){\line(1, -1){5}}
  \put(24, 7){$\phi_3$}
  \put(26, 2){$\phi_3$}
  \put(35, 4.5){$+$}
  \put(40, 4.5){\line(1, 0){6}}
  \put(40, 6){\line(1, 0){6}}
  {\color{red} \linethickness{1.5 pt}
    \qbezier(46, 6)(41.5, 11.5)(48, 11.5)
    \qbezier(48, 6)(44, 10)(48, 10)
    \qbezier(48, 6)(52, 10)(48, 10)
    \qbezier(50, 6)(54.5, 11.5)(48, 11.5)
  }
  \put(50, 4.5){\line(1, 0){5}}
  \put(50, 6){\line(1, 0){5}}
  \put(46, 4.5){\line(-1, -1){4}}
  \put(50, 4.5){\line(1, -1){4}}
  \put(48, 4.5){\line(1, -1){5}}
  \put(48, 4.5){\line(-1, -1){5}}
  \put(47, 2.5){$\phi_6$}
  {\color{white} 
    \multiput(24.5, 4.25)(1,0){3}{\circle*{.5}}
    \multiput(24.5, 5.75)(1,0){3}{\circle*{.5}}
  }
  \hspace*{-2.35mm}
  {\color{white}
    \put(47.1, 7){\circle*{0.6}}
    \put(45.2, 7){\circle*{0.6}}
    \put(48.9, 7){\circle*{0.6}}
    \put(50.8, 7){\circle*{0.6}}
    \put(46.4, 8){\circle*{0.6}}
    \put(44.6, 8){\circle*{0.6}}
    \put(49.6, 8){\circle*{0.6}}
    \put(51.4, 8){\circle*{0.6}}
    \put(46, 9){\circle*{0.6}}
    \put(44.2, 9){\circle*{0.6}}
    \put(50, 9){\circle*{0.6}}
    \put(51.8, 9){\circle*{0.6}}
    \put(47, 9.85){\circle*{0.6}}
    \put(44.55, 10.6){\circle*{0.6}}
    \put(48.8, 9.9){\circle*{0.6}}
    \put(51.4, 10.6){\circle*{0.6}}
    \put(50, 11.25){\circle*{0.6}}
    \put(46, 11.25){\circle*{0.6}}
    \put(48, 11.5){\circle*{0.6}}
   }
\end{picture}
\end{center}
\caption{\small An illustration of what the Polchinski equation does, using the double line notation appropriate for matrix fields \cite{Coleman}. The running of $\phi_4$ is caused by integrating out the red/dashed fast propagator in four-point diagrams that contain two $\phi_3$ vertices and one $\phi_6$ vertex with a loop. The double derivative of $e^{-S}$ in (\ref{PE}) gives the diagrams where integrations are to be performed: the second derivative of the action will give rise to one-loop diagrams and the square of the first derivative of the action will give rise to tree diagrams bridged by a fast mode propagator.}
\label{fig: PRG}
\end{figure}
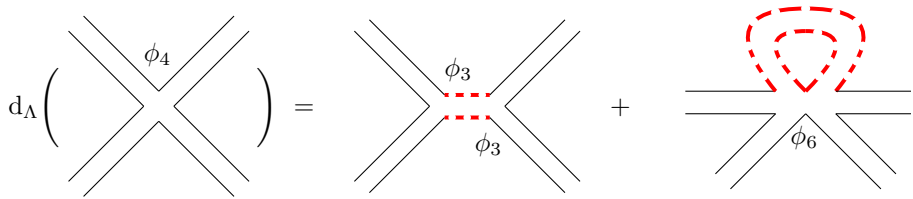

The PE looks unintuitive at first sight. All it does, however, is relate the actions before and after the RG step in terms of a simple integral over momenta in the momentum shell at cutoff $\Lambda$. This shell is infinitesimally thin, so the entire integral is proportional to $\d \Lambda$. The integrand consists of a high energy propagator that is being integrated out and of a second derivative of the exponential of the full action, which correctly captures all Feynman diagrams that contain a high energy propagator (cf.~Fig.~\ref{fig: PRG}).

The PE (\ref{PE}) also looks nothing like the holographic RG equation (\ref{flow}). As we have already suggested, we need to transform the action $S[M; \Lambda]$ in such a way so as to eliminate the explicit appearance of the fields $M$. It will transpire that a good idea is the use of a Hubbard-Stratonovich transformation (HST), suggested by \cite{Lee1} to connect bulk and boundary fields in a different way than presented here. The HST is typically used to get rid of an undesirable quadratic term in an exponential at the expense of introducing an extra integration; the idea is to exploit the relation
\bel{\label{HST}
  e^{\O^2/4a} = \sqrt{\frac a\pi}\int \d \varphi\ e^{- \varphi \O - a \varphi^2},
}
which introduces a new object $\varphi$ but also leaves only a linear dependence of the exponential on $\O$.\footnote{Practically (and typically), the Hubbard-Stratonovich transformation is used to evaluate the exponential on the l.h.s.~of (\ref{HST}) at small $a$, as then the r.h.s.~integral in (\ref{HST}) can be done using the saddle-point method. This approach does have some similarities with what we will do in our paper, as we will see.} However, the HST is simply the Laplace transformation of $e^{\O^2/4a}$, and hence we may also write a generalized transformation
\bel{
  e^{-f(\O)} = \int \d \varphi\ e^{- \varphi \O - \~f(\varphi)}.
}
It is clear that if the function $f$ also depends on some parameter other than $\O$, the transform $\~f$ will also depend on this parameter. This is why this transformation is useful for our present purposes. Such a transform of $S[M; \Lambda]$ would flow with $\Lambda$ but would depend only on the dummy variables $\varphi$ and the couplings $\phi$, but not on fields --- and we would then stand a chance of matching the two flow equations.

We thus propose a generalized Hubbard-Stratonovich transformation (GHST) by defining a transformed functional $\~S$ via a path integral over the dummy variables $\varphi$, which take all possible values of single-trace couplings:
\bel{\label{def S tilde}
  e^{-S[M; \Lambda]} \equiv \int [\d \varphi] e^{-N \varphi_n\cdot  \Tr M^n  - N^2 \~S [\varphi; \Lambda]}.
}
As this formula defines $e^{-N^2 \~S}$ as the Laplace transform of $e^{-S}$, the transformation from $S$ to $\~S$ is unique and formally invertible.  Notice that $\~S$ retains its dependence on $\Lambda$ and on all the couplings $\phi$ from the original action (\ref{def S}). In going from $S$ to $\~S$ we have replaced the dependence on a matrix field $M$ with the dependence on the scalar fields $\varphi$.

The $N^2$ prefactor in the exponent of the r.h.s.~above is inserted to make the $N$-counting work out in the rest of the paper, under the assumption that the leading order term in $\~S$ is of order $N^0$. This assumption will turn out to be self-consistent in eq.~(\ref{integral eq}). An additional perk is that this $N^2$ prefactor allows us to simply use the saddle-point method in the large $N$ limit, getting rid of the integral and reducing the GHST to a Legendre transformation. In this limit one recovers that the $\varphi$'s are set to be equal to single-trace couplings $\phi$. Because of this we will heuristically understand that, in general, the $\varphi$'s are mapped one-on-one onto single-trace couplings. Similar ideas related to the importance of single-trace couplings (and the use of the standard HST to generate double-trace couplings from single-trace ones) have already appeared in \cite{Heemskerk2}.

As already stressed, the immediate value of the GHST is that it separates out the $\Lambda$ dependence from the field content. The relevance of this becomes apparent after one glance at the PE (\ref{PE}), where the l.h.s.~is a manipulation of the cutoff $\Lambda$ and the r.h.s.~is a manipulation of the field content (with the propagator $\Delta(p)$ being thought of as ``the inverse of whatever is next to the term quadratic in the fields''). Thus, when we insert the GHST (\ref{def S tilde}) into (\ref{PE}), we get\footnote{Note that the exchange of the integral and the differential operator $\op H_\Lambda$ is possible because all functional integrals we work with are regularized --- i.e.~they are just multiple Riemannian integrals, and hence (under the standard assumption that all actions are convergent) we can interchange the order of integration and differentiation.}
\algnl{\label{PE2}
  &\dlam e^{-S} = \int [\d\varphi] e^{- N^2 \~S} \op H_\Lambda e^{-N \varphi_n \cdot \Tr M^n} .
}
The r.h.s.~contains a double derivative that can be written as an expression involving single- and double-trace terms:
\algnl{
   \op H_\Lambda e^{-N \varphi_n \cdot \Tr M^n} =- \frac{\d\Lambda}\Lambda\left(N\ \beta_n\cdot \Tr M^n + \frac12  \Tr M^n \cdot \gamma_{nm}\cdot \Tr M^m \right) e^{-N \varphi_n \cdot \Tr M^n}.
}
For the sake of brevity, we have introduced the parameters $\beta_n$ and $\gamma_{nm}$; their explicit forms will be given later in eqs.~(\ref{def beta}) and (\ref{def gamma}). The final, useful expression of how $\op H_\Lambda$ acts can now be retrieved by noting that
\bel{
  \Tr M^m e^{-N\varphi_n \cdot \Tr M^n} = - \frac1N \fder{}{\varphi_m} e^{-N\varphi_n \cdot \Tr M^n},
}
and so we can conclude that 
\algnl{
   \op H_\Lambda e^{- N \varphi_n \cdot \Tr M^n} 
   = \left(\beta_n \cdot \fder{}{\varphi_n} - \frac1{2N^2} \fder{}{\varphi_n} \cdot \gamma_{nm} \cdot \fder{}{\varphi_m} \right) e^{-N \varphi_n \cdot \Tr M^n}.
}

We can now insert this into the PE (\ref{PE2}) and integrate by parts. We assume that this can be done smoothly. Partial integration will give us terms such as $\fder{}{\varphi_n}(\beta_n e^{-N^2 \~S})$ and $\fder{^2}{\varphi_n\delta \varphi_m} (\gamma_{nm} e^{-N^2 \~S})$, but since both $\beta$'s and $\gamma$'s are of order $N^0$, the terms involving derivatives of $\beta_n$ and $\gamma_{nm}$ will be below leading order in $1/N$, and so they can be neglected. We thus get the flow equation
\algnl{\notag
   \dlam e^{-S}
   &= \int [\d\varphi]\  N^2\ \frac{\d\Lambda} \Lambda \left(\beta_n \cdot \fder{\~S}{\varphi_n}  - \frac12 \fder{\~S}{\varphi_n}\cdot \gamma_{nm} \cdot \fder{\~S}{\varphi_m} \right) e^{- N \varphi_n\cdot \Tr M^n  - N^2 \~S}\\  \label{integral eq}
   & = \int [\d \varphi] \left(- N^2 \dlam \~S \right) e^{- N \varphi_n\cdot \Tr M^n  - N^2 \~S}.
}
The fact that the Laplace transform (and therefore the GHST) is unique means that we can equate the integrands in the two integrals above without loss of generality; a solution $\~S$ that equates the integrands will necessarily equate the integrals, and this will then have to be the unique transform of the original action $S$. Writing everything in position space, this can be written as 
\bel{\label{flow S tilde}
  \Lambda\pder{\~S}\Lambda =  \int \xdvol \left[ \frac12 \gamma_{mn} \fder{\~S}{\varphi_n} \fder{\~S}{\varphi_m} - \beta_n \fder{\~S}{\varphi_n}\right].
}
Some care is needed to ensure that the functional derivatives are defined properly, due to the derivatives that are hidden within the couplings. The way to do it is to take all functional derivatives in momentum space and then to transform back to position space. 

The flow equation (\ref{flow S tilde}) indeed resembles the holographic RG equation (\ref{flow}), but more manipulations need to be done to show that these are actually the same equation. The first thing to do is to isolate the explicit cutoff dependence from $\beta_n$ and $\gamma_{nm}$. These parameters can be calculated by straightforward double differentiation of $e^{-N\varphi_n \cdot \Tr M^n}$ found on the r.h.s.~of eq.~(\ref{PE2}), which yields (again, to leading order in $1/N$)
\algnl{\notag
  \beta_n \frac{\d\Lambda}\Lambda
  &\equiv \int_{\Lambda - \d\Lambda}^\Lambda \pdnvol \frac{1}{2\varphi_2(p)} \Biggr[ - (n + 2) \varphi_{n + 2}(p, -p, p_1, \ldots, p_n) + \\ \label{def beta}
  &+ \sum_{m = 2}^{n - 2}  (m + 1) (n - m + 1) \varphi_{m + 1}(p, p_1, \ldots, p_m) \varphi_{n - m + 1}(-p, p_{m + 1}, \ldots, p_n) \Biggr]
}
and
\algnl{\label{def gamma}
  \gamma_{nm}\frac{\d\Lambda}\Lambda 
  &\equiv -\int_{\Lambda - \d\Lambda}^\Lambda \pdnvol \frac{1}{2\varphi_2(p)} (n + m + 2) \varphi_{m + n + 2}(p, p_1, \ldots, p_n, -p, q_1, \ldots, q_m).
}
Note that instead of writing $\dlam K_\Lambda(p^2)$ everywhere, we indicate that we are integrating over the momentum shell at $\Lambda$ by writing $\int_{\Lambda - \d\Lambda}^\Lambda$. 

The parameters $\beta_n$ and $\gamma_{nm}$ are basically the $\beta$-functions of single- and double-trace couplings. In fact, using the flow equation (\ref{flow S tilde}) for $\~S$ one can indeed show that, at $N \gg 1$,
\bel{
  \Lambda \der{\phi_n}{\Lambda} = \beta_n \big|_{\varphi = \phi}.
}
Fig.~\ref{fig: PRG} precisely illustrates how the formula for $\beta$ works out. Two kinds of Feynman diagrams contribute as we integrate out the fast propagator from diagrams which contain such a  propagator. Both kinds are illustrated on the r.h.s.~of this figure. On the one hand, one-particle reducible tree-level diagrams (like the one with two cubic vertices) can get joined together, yielding corrections to couplings that are proportional to products of two other couplings. On the other hand, loops (like the one with the six-point vertex) can get contracted to yield corrections that are proportional to only one other coupling. The corrections to double-trace couplings, represented by $\gamma$, can stem only from loops getting contracted, and hence these corrections are proportional to only one of the other couplings. Some further discussion of the objects $\beta_n$ and $\gamma_{nm}$ can also be found in section 9.4 of \cite{Banks}, where they are developed in the context of the Polchinski RG equation for a scalar QFT.

Going back to the definitions (\ref{def beta}) and (\ref{def gamma}), we can use the assumption that the couplings are small, and write, in momentum space,
\bel{
  \beta_n \frac{\d\Lambda}\Lambda = -\int_{\Lambda - \d\Lambda}^\Lambda \pdnvol \frac{1}{2\varphi_2(p)} (n + 2) \varphi_{n + 2}(p, -p, p_1, \ldots, p_n).
}
Both $\beta_n$ and $\gamma_{nm}$ are now given as  momentum shell integrals of a ratio of two couplings ($\varphi_{n + 2}/\varphi_2$ and $\varphi_{n + m + 2}/\varphi_2$, respectively). Expanding couplings in powers of momenta will yield finite-degree polynomials, as per our assumption that couplings should contain only a finite number of derivatives. One momentum that enters the argument (labeled $p$ in (\ref{def beta}) and (\ref{def gamma})) will take values close to $\Lambda$. If this term is thought to dominate the other terms and if no coupling contains higher powers of momenta than $\varphi_2$, we can then notice that the ratios of couplings will, to leading order, behave as $\Lambda^0$, and so $\beta_n$ and $\gamma_{nm}$ will both scale as the momentum shell volume $\Lambda^d$ with additional corrections in powers below $\Lambda^d$. 

We now note that we can ``complete the square''  in eq.~(\ref{flow S tilde}). If we assume that we can choose a functional $W[\varphi]$ such that $\beta$ can be written as a gradient flow,  then we can define the shifted action and the appropriate conjugate momenta:
\bel{
  \beta_n \equiv \gamma_{nm}\cdot \fder W{\varphi_m},\quad  I[\varphi; \Lambda] \equiv \~S[\varphi; \Lambda] - W[\varphi], \quad \pi^n \equiv \fder{I}{\varphi_n}.
}
With these definitions, by extracting the $\Lambda^d$ scaling from $\beta_n$ and $\gamma_{mn}$ (along with an arbitrary constant $R$), we can introduce the new quantities
\bel{
  \eps \equiv \frac1\Lambda,\quad G_{nm} \equiv - \frac{\eps^d}{R^{d + 1}} \gamma_{nm},\quad V(\varphi, \eps) \equiv \frac12 G_{nm} \fder{W}{\varphi_n}\cdot \fder W{\varphi_m}.
}
Both $V$ and $G_{nm}$ scale as $\eps^0$ to leading order in $\eps$, as in eqs.~(\ref{V}) and (\ref{flow}) for the holographic RG. Moreover, the flow equation (\ref{flow S tilde}) now takes the form
\bel{\label{flow I}
  \eps \pder I\eps = R \int \d^d x\ \frac{R^d}{\eps^d}  \left[\frac12 G_{nm} \pi^n\pi^m - V(\varphi, \eps)\right].
}
This is formally the same equation as the one coming from the holographic RG (\ref{flow}). Notice, in particular, that the potential $V$ coming from the boundary will have corrections of order $\eps^2$ as long as the propagator has at least a dependence on $p^2$ and a subleading mass term, as in the standard case. It is reassuring that these corrections naturally arise in the bulk theory as kinetic terms on the bulk boundary, as can be seen in eq.~(\ref{V}).

\section{Outlook}

We have explicitly demonstrated how the holographic and Wilsonian RGs lead to the same flow equation. A classical scalar field theory in an anti-de Sitter space can thus be dual to a global $O(N)$ quantum theory of a single matrix field in the large $N$ limit. The fact that we have managed to mimic the AdS/CFT correspondence so closely gives us hope that considerations such as ours might be used to derive more aspects of string theory from deep properties of quantum field theories.

In principle, the equivalence of eqs.~(\ref{flow}) and (\ref{flow I}) allows us to write some form of an AdS/CFT dictionary from ``scratch,'' ignoring string theory completely. However, this is not so simple at present. The most obvious hurdle is that going from bulk to boundary (or vice versa) requires inverting $G_{nm}$ or $\gamma_{nm}$, and this is difficult because we are working with an infinite number of couplings on each side. It seems that we should be able to develop a truncation procedure, where we can integrate out (or discard altogether) the majority of the fields on both sides; then we should be able to diagonalize these matrices and easily find their inverses. This is a plausible proposition because the potential $V$ in (\ref{flow I}) can be rewritten as
\bel{
  V = \frac12 \left(\frac{\eps^d}{R^{d+1}}\right)^2 G^{nm} \beta_n \beta_m.
}
This potential clearly depends on the square of $\beta$-functions of boundary couplings, and hence couplings with large scaling dimensions (and with $\beta_n \sim \Delta_{nm} \phi_m$) will have potentials $V \sim \Delta_n^2 \phi_n^2$. In the bulk, large $\Delta_n$ would correspond to very massive fields; in the boundary, it would correspond to highly irrelevant operators. In either case, we expect these fields to effectively remain static. This seems connected to the notion that  locality of the bulk potential is related to gaps in the operator spectrum on the boundary \cite{Heemskerk1}. This idea and our present work suggest that tangible potentials will be achieved only when we can formulate a truncation procedure for operators. This remains a topic for future work.

A number of points also warrant further clarification, and again these will all be deferred to future work. It remains unclear what would happen if $\beta_n$ were to be kept non-linear in couplings; this would signify moving away from the fixed point, and this should correspond to moving away from the AdS metric in the bulk. It would also be valuable to go beyond the leading order in $1/N$, especially if we were to try to tackle realistic theories which probably do not have large $N$. We have outlined places where $1/N$ would enter; the AdS/CFT dictionary suggests that these corrections should correspond to loop corrections in the bulk, and it would be interesting to see whether further progress can be made on this front. In an effort to reach out towards realistic theories, it would also be desirable to attempt this same argument for a more complicated boundary theory --- perhaps one with spinor or vector fields as well, or one with a local gauge symmetry (in which case a hard cutoff is no longer the gauge-invariant approach, and new ideas are needed). It would also be interesting to study the relevance of the AdS radius as a measure of nonlocality in the bulk, as this concept has never figured significantly in this paper. Another quantity that deserves further thought is the function $W$; it would be nice to understand its locality and, perhaps, to show that it exists in the first place. Finally, it would be interesting to study the stability of our result under Wick rotations, as this would allow us to extend the discussion to real time.

\section{Acknowledgments}

This work has been done as an undergraduate senior thesis project at Princeton University. The author would like to thank his advisor, Herman Verlinde, for the many insights and constant encouragement throughout the entire project. The author would also like to express his appreciation to the anonymous referee of this article, whose detailed reading and many sharp comments made this manuscript much more lucid.

\bibliographystyle{apsrev}
%\bibliography{../Props/AdS-CFT}
%\newpage
\bibliography{Holography_and_RG_arXiv_v4.bbl}
\end{document}